\date{}
\documentstyle [11pt]{article}
\newtheorem{theo}{Theorem}[section]

\newtheorem{lemma}[theo]{Lemma}
\newtheorem{coro}[theo]{Corollary}

\renewcommand{\baselinestretch}{1.5}

\begin{document}
\thispagestyle{empty}
\begin{center}
TEL-AVIV UNIVERSITY\\
RAYMOND AND BEVERLY SACKLER\\
FACULTY OF EXACT SCIENCES\\
\vspace{30 mm}

{\huge
Choice numbers of graphs
}

\vspace{15 mm}
This work was submitted in partial fulfillment of the \\
requirements for the Master's degree (M.Sc.) at \\
Tel-Aviv University \\
\vspace{15 mm}
School of Mathematical Sciences\\
Department of Computer Science\\
\vspace{15 mm}
Presented by\\
Shai Gutner\\
\vspace{20 mm}
This work was carried out under the supervision of \\
Prof. Michael Tarsi\\
\vspace{10 mm}
July 1992
\end{center}
\newpage
\noindent
{\em I am grateful to my advisor,}

\noindent
{\em Prof. Michael Tarsi, for his}

\noindent
{\em guidance and assistance}

\noindent
{\em throughout the work on this}

\noindent
{\em thesis.}

\vspace{10 mm}

\noindent
{\em I would like to thank}

\noindent
{\em Prof. Noga Alon for helpful}

\noindent
{\em discussions and comments.}

\newpage

\tableofcontents

\newpage
\section{Introduction}

A graph $G=(V,E)$ is {\em $(a:b)$-choosable} if for every family of 
sets $\{S(v): v \in V \}$, where $|S(v)|=a$ for all $v \in V$, there are 
subsets $C(v) \subseteq S(v)$, where $|C(v)|=b$ for all $v \in V$, and 
$C(u) \cap C(v)= \emptyset$ for every two adjacent vertices $u,v \in V$.
The {\em $k$th choice number} of $G$, denoted by $ch_k(G)$,
is the minimum integer $n$ so that $G$ is $(n:k)$-choosable.
A graph $G=(V,E)$ is {\em $k$-choosable} if it is $(k:1)$-choosable.
The {\em choice number} of $G$, denoted by $ch(G)$, is equal to $ch_1(G)$.

The concept of $(a:b)$-choosability was defined and studied by Erd\H{o}s,
Rubin and Taylor in~\cite{ERT}. In the present paper we prove several results
concerning $(a:b)$-choosability, a number of which generalize known
results regarding choice numbers of graphs that appear in~\cite{AT} and
\cite{A2}. The following theorem examines the behavior of $ch_k(G)$ when
$k$ is large.
\begin{theo}
\label{t21}
Let G be a graph. For every $\epsilon >0$ there exists an integer $k_0$ such 
that $ch_k(G) \leq k(\chi (G)+\epsilon)$ for every $k \geq k_0$.
\end{theo}
In~\cite{ERT} the authors ask the following question: \\
If $G$ is $(a:b)$-choosable, and $\frac{c}{d} > \frac{a}{b}$, does it 
follow that G is $(c:d)$-choosable? \\
The following corollary gives a negative answer to this question.
\begin{coro}
\label{c21}
If $l > m \geq 3$, then there is a graph $G$ which is $(a:b)$-choosable
but not $(c:d)$-choosable where $\frac{c}{d} = l$ and $\frac{a}{b} = m$.
\end{coro}

Let $K_{m*r}$ denote the complete $r$-partite graph with $m$ vertices in
each vertex class, and let $K_{m_1, \ldots ,m_r}$ denote the complete
$r$-partite graph with $m_i$ vertices in the $i$th vertex class.
It is shown in~\cite{A2} that there exist two positive constants $c_1$
and $c_2$ such that for every $m \geq 2$ and for every $r \geq 2$,
$c_1 r \log m \leq ch(K_{m*r}) \leq c_2 r \log m$. The following theorem
generalizes the upper bound.
\begin{theo}
\label{t31}
If $r \geq 1$ and $m_i \geq 2$ for every $i$, $1 \leq i \leq r$, then
$$
ch_k(K_{m_1, \ldots ,m_r}) \leq 948 r(k+\log{\frac{m_1+ \cdots + m_r}{r}}).
$$
\end{theo}
The following are two applications of this theorem.
\begin{coro}
\label{c31}
For every graph G and $k \geq 1$
$$
ch_k(G) \leq 948  \chi (G) (k+\log{(\frac{|V|}{\chi(G)}+1)}).
$$
\end{coro}
The second corollary generalizes a result from~\cite{A2} concerning the
choice numbers of random graphs for the common model $G_{n,p}$
(see, e.g.,~\cite{Bo2}), in which the graph is obtained by taking each
pair of the $n$ labeled vertices $1,2, \ldots ,n$ to be an edge, 
randomly and independently, with probability $p$.  
\begin{coro}
\label{c32}
For every two constants $k \geq 1$ and $0 < p < 1$, the probability that 
$ch_k(G_{n,p}) \leq 475 \log{(1/(1-p))} n \frac{\log \log n}{\log n}$ tends 
to $1$ as $n$ tends to infinity.
\end{coro}

A theorem which appears in~\cite{AT} reveals the connection between the
choice number of a graph $G$ and its orientations. We present here a
generalization of this theorem for a special case.
\begin{theo}
\label{t41}
Let $D=(V,E)$ be a digraph and $k \geq 1$. For each $v \in V$, let $S(v)$
be a set of size $k(d^+_D(v)+1)$, where $d^+_D(v)$ is the outdegree of v.
If D contains no odd directed (simple) cycle, then there are subsets 
$C(v) \subseteq S(v)$, where $|C(v)|=k$ for all $v \in V$, and 
$C(u) \cap C(v)= \emptyset$ for every two adjacent vertices $u,v \in V$.
There is a polynomial time algorithm in $|V|$ and $k$ which finds the 
subsets $C(v)$.
\end{theo}
\begin{coro}
\label{c41}
Let G be an undirected graph. If $G$ has an orientation $D$ which contains 
no odd directed (simple) cycle in which the maximum outdegree is d, then
$G$ is $(k(d+1):k)$-choosable for every $k \geq 1$.
\end{coro}
\begin{coro}
\label{c42}
An even cycle is $(2k:k)$-choosable for every $k \geq 1$.
\end{coro}
The last corollary enables us to prove a generalization of a variant of
Brooks Theorem which appears in~\cite{ERT}.
\begin{coro}
\label{c43}
If a connected graph $G$ is not $K_n$, and not an odd cycle,
then $ch_k(G) \leq k \Delta(G)$ for every $k \geq 1$, where $\Delta(G)$
is the maximum degree of $G$.
\end{coro}
For a graph $G=(V,E)$, define $M(G)=\max(|E(H)|/|V(H)|)$,
where $H=(V(H),E(H))$ ranges over all subgraphs of $G$. The
following two corollaries are generalizations of results which appear
in~\cite{AT}.
\begin{coro}
\label{c44}
Every bipartite graph $G$ is
$(k(\lceil M(G)\rceil+1):k)$-choosable 
for all $k \geq 1$.
\end{coro}
\begin{coro}
\label{c45}
Every bipartite planar graph $G$ is $(3k:k)$-choosable for 
all $k \geq 1$.            
\end{coro}
The following are additional applications.
\begin{coro}
\label{c46}
If every induced subgraph of a graph $G$ has a vertex of degree at most $d$,
then $G$ is $(k(d+1):k)$-choosable for all $k \geq 1$.
\end{coro}
\begin{coro}
\label{c47}
If $G$ is a triangulated graph, then $ch_k(G)=k\chi (G)=k\omega (G)$ for 
every $k \geq 1$, where $\omega (G)$ is the clique number of $G$.
\end{coro}

The list-chromatic conjecture asserts that for every graph $G$,
$ch(L(G))=\chi (L(G))$, where $L(G)$ is the line graph of $G$.
The list-chromatic conjecture is easy to establish for trees, graphs of       
degree at most 2, and $K_{2,m}$. It has also been verified for
snarks~\cite{H}, $K_{3,3}$, $K_{4,4}$, $K_{6,6}$~\cite{AT}, and 2-connected
cubic planar graphs. The following corollary shows that the list-chromatic
conjecture is true for graphs which contain no $C_n$ for 
every $n \geq 4$.
\begin{coro}
\label{c48}
If a graph $G$ contains no $C_n$ for every $n \geq 4$, then
$ch(L(G))=\chi (L(G))$.
\end{coro}

The {\em core} of a graph $G$ is the graph obtained from $G$ by deleting
nodes of degree $1$ successively until there are no nodes of degree $1$.
The graph $\Theta_{a,b,c}$ consists of two distinguished nodes $u$ and $v$
together with three paths of lengths $a$,$b$, and $c$, which are node 
disjoint except that each path has $u$ at one end, and $v$ at the other end. 
The following theorem from~\cite{ERT} gives a characterization of the
$2$-choosable graph:
\begin{theo}
\label{t51}
A connected graph $G$ is $2$-choosable if, and only if, the core of $G$ 
belongs to $\{ K_1 , C_{2m+2} , \Theta_{2,2,2m} : m \geq 1 \}$.
\end{theo} 
In~\cite{ERT} the authors ask the following question: \\
If $G$ is $(a:b)$-choosable, does it follow that $G$ is $(am:bm)$-choosable? \\
The following theorem gives a partial solution to this question by using
theorem~\ref{t51}.
\begin{theo}
\label{t52}
If a graph $G$ is $2$-choosable, then $G$ is also $(4:2)$-choosable.
\end{theo}
\begin{theo}
\label{t53}
Suppose that $k$ and $m$ are positive integers and that $k$ is odd. If a graph 
$G$ is $(2mk:mk)$-choosable, then $G$ is also $2m$-choosable.
\end{theo}

A graph $G=(V,E)$ is {\em $f$-choosable} for a function $f: V \mapsto N$     
if for every family of sets $\{S(v): v \in V \}$, where $|S(v)|=f(v)$ for   
all $v \in V$, there is a proper vertex-coloring of $G$ assigning to
each vertex $v \in V$ a color from $S(v)$.
It is shown in~\cite{ERT} that the following problem is
$\Pi _2^p$-complete: ( for terminology see~\cite{GJ} )

\vspace{3 mm}
\noindent
{\bf BIPARTITE GRAPH $(2,3)$-CHOOSABILITY (BG $(2,3)$-CH)} \\
INSTANCE: A bipartite graph $G=(V,E)$ and a 
function $f: V \mapsto \{2,3\}$. \\
QUESTION: Is $G$ $f$-choosable? 

\vspace{3 mm}
\noindent
We consider the following decision problem:

\vspace{3 mm}
\noindent
{\bf BIPARTITE GRAPH $k$-CHOOSABILITY (BG $k$-CH)} \\
INSTANCE: A bipartite graph $G=(V,E)$. \\
QUESTION: Is $G$ $k$-choosable? 

\vspace{3 mm}
\noindent
If follows from theorem~\ref{t51} that this problem is solvable in
polynomial time for $k=2$. 
\begin{theo}
\label{t61}
{\bf BIPARTITE GRAPH $k$-CHOOSABILITY} is $\Pi _2^p$-complete for
every constant $k \geq 3$.
\end{theo}  

A graph $G=(V,E)$ is {\em strongly $k$-colorable} if every graph obtained
from $G$ by adding to it a union of vertex disjoint cliques of size at     
most $k$ ( on the set $V$ ) is $k$-colorable. 
An analogous definition of {\em strongly $k$-choosable} is made by replacing
colorability with choosability.
The {\em strong chromatic number} of a graph $G$, denoted by $s\chi(G)$, is
the minimum $k$ such that $G$ is strongly $k$-colorable. 
Define $s\chi(d) = \max(s\chi(G))$, where $G$ ranges over all graphs with
maximum degree at most $d$. 
The definition of strongly $k$-colorable given in~\cite{A1} is
slightly different. It is claimed there that if $G$ is strongly $k$-colorable,
then it is strongly $(k+1)$-colorable as well. However, it is not known
how to prove this if we use the definition from~\cite{A1}.
\begin{theo}
\label{t71}
If $G$ is strongly $k$-colorable, then it is strongly $(k+1)$-colorable
as well.
\end{theo}
We give a weaker version of this theorem for choosability.
\begin{theo}
\label{t72}
If $G$ is strongly $k$-choosable, then it is strongly $km$-choosable as well.
\end{theo}
\begin{theo}
\label{t73}
Let $G=(V,E)$ be a graph, and suppose that $km$ divides $|V|$. If the
choice number of any graph obtained from $G$ by adding to it a union of
vertex disjoint $k$-cliques (on the set $V$) is $k$, then the choice number
of any graph obtained from $G$ by adding to it a union of
vertex disjoint $km$-cliques is $km$.
\end{theo}
\begin{coro}
\label{c71}
Let $n$ and $k$ be positive integers, and let $G$ be a $(3k+1)$-regular graph
on $3kn$ vertices. Assume that $G$ has a decomposition into a Hamiltonian
circuit and $n$ pairwise vertex disjoint $3k$-cliques. Then $ch(G)=3k$.
\end{coro}

It is proved in~\cite{A1} that there is a constant $c$ such that for
every $d$, $3 \lfloor d/2 \rfloor < s\chi(d) \leq cd$. The following
theorem improves the lower bound.
\begin{theo}
\label{t74}
For every $d \geq 1$, $s\chi(d) \geq 2d$.
\end{theo}

\section{A solution to a problem of Erd\H{o}s, Rubin and Taylor}

In this section we prove an upper bound for the $k$th choice number of
a graph when $k$ is large and apply this bound to settle a problem
raised in~\cite{ERT}.

\noindent
{\bf Proof of Theorem~\ref{t21}}\,
Let $G=(V,E)$ be a graph and $\epsilon >0$. Denote $r=\chi (G)$, and let 
$V=V_1 \cup \ldots \cup V_r$ be a partition of the vertices, such that 
each $V_i$ is a stable set. For each $v \in V$, let $S(v)$ be a set of 
$\lfloor k(\chi (G)+\epsilon) \rfloor$ distinct colors. Let 
$S=\cup_{v \in V} S(v)$ be the set of all colors. Put
$R=\{1,2, \ldots ,r \}$ and let $f: S \mapsto R$ be a random function,
obtained by choosing, for each color $c \in S$, randomly and independently,   
the value of $f(c)$ according to a uniform distribution on $R$. The colors
$c$ for which $f(c)=i$ will be the ones to be used for coloring the 
vertices in $V_i$. To complete the proof, it thus suffices to
show that with positive probability for every $i$, $1 \leq i \leq r$, and
for every vertex $v \in V_i$ there are at least $k$ colors $c \in S(v)$
so that $f(c)=i$.

Fix an $i$ and a vertex $v \in V_i$, and define $X=|S(v) \cap f^{-1}(i)|$.
The probability that there are less than $k$ colors $c \in S(v)$ so
that $f(c)=i$ is equal to $Pr(X<k)$. Since $X$ is a random variable with 
distribution $B(\lfloor k(r+\epsilon ) \rfloor  ,1/r)$, by Chebyshev's 
inequality (see, e.g.,~\cite{AS})
$$
Pr(X<k) \leq 
Pr(| X-\frac{\lfloor k(r+\epsilon) \rfloor }{r} | \geq
\frac{\lfloor k\epsilon \rfloor }{r} ) \leq
\frac{\lfloor k(r+\epsilon ) \rfloor \frac{1}{r} (1-\frac{1}{r})}
{(\frac{\lfloor k\epsilon \rfloor }{r})^2} =  O(\frac{1}{k}).
$$
It follows that there is an integer $k_0$ such that $P(X<k)<1/|V|$ for every 
$k \geq k_0$. There are $|V|$ possible choices of $i$, $1 \leq i \leq r$ 
and $v \in V_i$, and hence, the probability that for some $i$ and some 
$v \in V_i$ there are less than $k$ colors $c \in S(v)$ so that $f(c)=i$
is smaller than $1$, completing the proof. $\Box$

Note that it is not true that for every graph $G$ there exists an integer
$k_0$ such that $ch_k(G) \leq k \chi(G)$ for every $k \geq k_0$.
For example, take the graph $G=K_{3,3}$ which has chromatic number $2$.
The graph $G$ is not $2$-choosable and therefore by theorem~\ref{t53} 
it is not $(2k:k)$-choosable for every $k$ odd. This means that 
$ch_k(G) > k \chi(G)$ for every $k$ odd. 

\noindent
{\bf Proof of Corollary~\ref{c21}}\,
Suppose that $l > m \geq 3$, and let G be a graph such that $ch(G)=l+1$
and $\chi (G)=m-1$ ( it is proved in~\cite{V} that for every
$l \geq m \geq 2$ there is a graph $G$, where $ch(G)=l$ and $\chi (G)=m$ ). 
By theorem~\ref{t21}, for $\epsilon =1$ there exist an integer $k$ such
that $G$ is $(k(\chi(G)+1):k)$-choosable. We have that $G$ is
$(km:k)$-choosable but not $(l:1)$-choosable, as needed. $\Box$

\section{An upper bound for the $k$th choice number}

In this section we establish an upper bound for 
$ch_k(K_{m_1, \ldots ,m_r})$, and use it to prove two consequences.
The following lemma appears in~\cite{AS}.
\begin{lemma}
\label{l31}
If $X$ is a random variable with distribution $B(n,p)$, $0 < p \leq 1$,
and $k<pn$ then
$$
Pr(X<k) < e^{-\frac{(np-k)^2}{2pn}}.
$$
\end{lemma}

In the rest of this section we denote $t=\frac{m_1+ \cdots +m_r}{r}$,
$t_1=\frac{m_1+ \cdots +m_{r/2}}{r/2}$, and 
$t_2=\frac{m_{r/2+1}+ \cdots +m_r}{r/2}$.
Notice that $t=(t_1+t_2)/2$, and therefore $\log{t_1 t_2} \leq 2\log{t}$. 
\begin{lemma}
\label{l32}
If $1 \leq r \leq t$, $k \geq 1$, and $m_i \geq 2$ for every $i$,
$1 \leq i \leq r$, then    
$ch_k(K_{m_1, \ldots ,m_r}) \leq 4r(k+\log{t})$.
\end{lemma}
{\bf Proof}\,
Let $V_1, V_2, \ldots ,V_r$ be the vertex classes of
$K = K_{m_1, \ldots ,m_r}$, where $|V_i|=m_i$ for all $i$, and let
$V = V_1 \cup \ldots \cup V_r$ be the set of all vertices of $K$.
For each $v \in V$, let $S(v)$ be a set of $\lfloor 4r(k+\log{t}) \rfloor$
distinct colors. Put $R=\{1,2, \ldots ,r \}$ and let $f: S \mapsto R$ be 
a random function, obtained by choosing, for each color $c \in S$, randomly
and independently, the value of $f(c)$ according to a uniform distribution  
on $R$. The colors $c$ for which $f(c)=i$ will be the ones to be used for   
coloring the vertices in $V_i$. To complete the proof it thus suffices to
show that with positive probability for every $i$, $1 \leq i \leq r$,
and every vertex $v \in V_i$ there are at least $k$ colors $c \in S(v)$
so that $f(c)=i$.

Fix an $i$ and a vertex $v \in V_i$, and define $X=|S(v) \cap f^{-1}(i)|$.
The probability that there are less than $k$ colors $c \in S(v)$ so
that $f(c)=i$ is equal to $Pr(X<k)$. Since $X$ is a random variable with
distribution $B(\lfloor 4r(k+\log{t}) \rfloor,1/r)$, by lemma~\ref{l31}
$$
Pr(X<k) < e^{-\frac{(E(X)-k)^2}{2E(X)}} \leq
e^{-\frac{(4(k+\log{t})-1-k)^2}{8(k+\log{t})}} <
e^{-\frac{16(k+\log{t})^2-8(k+1)(k+\log{t})}{8(k+\log{t})}} \leq 
e^{-2\log{t}} = \frac{1}{t^2} \leq \frac{1}{rt},  
$$
where the last inequality follows from the fact that $r \leq t$. 
There are $rt$ possible choices of $i$, $1 \leq i \leq r$ and $v \in V_i$,
and hence, the probability that for some $i$ and some $v \in V_i$ there
are less than $k$ colors $c \in S(v)$ so that $f(c)=i$ is smaller than $1$,
completing the proof. $\Box$

\begin{lemma}
\label{l33}
Suppose that $r$ is even, $r > t$, $k \geq 1$, $d \geq 244$, and
$m_i \geq 2$ for every $i$, $1 \leq i \leq r$. If 
$ch_k(K_{m_1, \ldots ,m_{r/2}}) \leq 
d(1-\frac{1}{5 r^{1/3}})\frac{r}{2}(k+ \log{t_1})$ and  
$ch_k(K_{m_{r/2+1}, \ldots ,m_r}) \leq 
d(1-\frac{1}{5 r^{1/3}})\frac{r}{2}(k+ \log{t_2})$, then
$ch_k(K_{m_1, \ldots ,m_r}) \leq dr(k+ \log{t})$.
\end{lemma}
{\bf Proof}\,
Let $V_1, V_2, \ldots ,V_r$ be the vertex classes of 
$K=K_{m_1, \ldots ,m_r}$, where $|V_i|=m_i$ for all $i$, and let
$V=V_1 \cup \ldots \cup V_r$ be the set of all vertices of $K$.
For each $v \in V$, let $S(v)$ be a set of $\lfloor dr(k+ \log{t}) \rfloor$
distinct colors. Define $R=\{1,2, \ldots ,r\}$, and let 
$S=\cup_{v \in V} S(v)$ be the set of all colors. Put 
$R_1=\{1,2, \dots ,r/2\}$ and $R_2=\{r/2+1, \ldots ,r\}$.    
Let $f:S \mapsto \{1,2\}$ be a random function obtained by choosing, for
each $c \in S$ randomly and independently, $f(c) \in \{1,2\}$ where for
all $j \in \{1,2\}$ 
$$
Pr(f(c)=j)=\frac{k+\log{t_j}}{2k+\log t_1 t_2}.
$$ 
The colors $c$ for which $f(c)=1$ will be used for coloring the vertices
in $\cup_{i \in R_1} V_i$, whereas the colors $c$ for which $f(c)=2$
will be used for coloring the vertices in $\cup_{i \in R_2} V_i$.

For every vertex $v \in V$, define $C(v)=S(v) \cap f^{-1}(1)$ if $v$
belongs to $\cup_{i \in R_1} V_i$, and $C(v)=S(v) \cap f^{-1}(2)$
if $v$ belongs to $\cup_{i \in R_2} V_i$. Because of the assumptions of 
the lemma, it remains to show that with positive probability,
\begin{equation}
\label{e31}
|C(v)| \geq d(1-\frac{1}{5 r^{1/3}})\frac{r}{2}(k+\log{t_j})
\end{equation}
for all $j \in \{1,2\}$ and $v \in \cup_{i \in R_j} V_i$.

Fix a $j \in \{1,2\}$ and a vertex $v \in \cup_{i \in R_j} V_i$, and 
define $X=|C(v)|$. The expectation of $X$ is
$$
\lfloor dr(k+\log{t}) \rfloor \frac{k+\log{t_j}}{2k+\log t_1 t_2} \geq
(dr(k+\log{t}) -1) \frac{k+\log{t_j}}{2k+2\log{t}} \geq
d\frac{r}{2}(k+\log{t_j}) -1 = T.
$$
If follows from lemma~\ref{l31} and the inequality $E(X) \geq T$ that
$$
Pr(X<T-T^{2/3})<e^{-\frac{(E(X)-T+T^{2/3})^2}{2E(X)}} \leq
e^{-\frac{1}{2} T^{1/3}} \leq e^{-\frac{1}{2}(d\frac{r}{2})^{1/3}}.
$$
Since $|\cup_{i \in R_j} V_i| \leq rt < r^2$, the probability that 
$|C(v)|<T-T^{2/3}$ holds for some $v \in \cup_{i \in R_j} V_i$ 
is at most
$$
r^2 \cdot e^{-\frac{1}{2}(d\frac{r}{2})^{1/3}} < 1/2 ,
$$
where the last inequality follows from the fact that $d \geq 244$.
One can easily check that
$$
T-T^{2/3}=T(1-\frac{1}{T^{1/3}}) \geq 
d\frac{r}{2}(k+\log{t_j})(1-\frac{1}{5 r^{1/3}}),
$$
and therefore, with positive probability (\ref{e31}) holds for all 
$j \in \{1,2\}$ and $v \in \cup_{i \in R_j} V_i$. $\Box$

\noindent
{\bf Proof of Theorem~\ref{t31}}\,
Define for every $r$ which is a power of $2$	
$$
f(r)=\prod_{j=0}^{\log_2 r} (1-\frac{1}{5 \cdot 2^{j/3}}) /
\prod_{j=0}^{2} (1-\frac{1}{5 \cdot 2^{j/3}}).
$$
We claim that for every $r$ which is a power of $2$
\begin{equation}
\label{e32}
ch_k(K_{m_1, \ldots ,m_r}) \leq \frac{244r(k+\log{t})}{f(r)}.
\end{equation}
The proof is by induction on $r$.

\noindent
{\bf Case 1:}\, $r \leq t$.

\noindent
The result follows from lemma~\ref{l32} since
$$
\frac{244}{f(r)} \geq
244 \prod_{j=1}^{2} (1-\frac{1}{5 \cdot 2^{j/3}}) > 4.
$$

\noindent
{\bf Case 2:}\, $r > t$.

\noindent 
Notice that $t \geq 2$, and therefore $r \geq 4$. By the induction hypothesis 
$$
ch_k(K_{m_1, \ldots ,m_{r/2}}) \leq 
\frac{244(1-\frac{1}{5 r^{1/3}})\frac{r}{2}(k+\log{t_1})}{f(r)}
$$
and
$$
ch_k(K_{m_{r/2+1}, \ldots ,m_r}) \leq
\frac{244(1-\frac{1}{5 r^{1/3}})\frac{r}{2}(k+\log{t_2})}{f(r)}.
$$
Since $r \geq 4$, we have $244/f(r) \geq 244$ and it follows from lemma
\ref{l33} that (\ref{e32}) holds, as claimed.

It is easy to check that
$$
\prod_{j=3}^{\log_2 r} (1-\frac{1}{5 \cdot 2^{j/3}}) \geq
1 - \sum_{j=3}^{\log_2 r} \frac{1}{5 \cdot 2^{j/3}} \geq
1 - \frac{1}{10(1 - 2^{-1/3})},    
$$
and therefore $244/f(r) \leq 474$. If follows from (\ref{e32}) that
for every $r$ which is a power of $2$
\begin{equation}
\label{e33}
ch_k(K_{m_1, \ldots ,m_r}) \leq 474r(k+\log{t}).
\end{equation}

Returning to the general case, assume that $r \geq 1$. Choose an integer
$r'$ which is a power of $2$ and $r < r' \leq 2r$. By applying
(\ref{e33}),  we get
$$
ch_k(K_{m_1, \ldots ,m_r}) \leq
ch_k(K_{m_1, \ldots ,m_r, \underbrace{2, \ldots ,2}_{r'-r}})
$$
$$
\leq 474 r'(k+\log{\frac{m_1 + \cdots + m_r + 2(r'-r)}{r'}}) \leq
948 r(k+\log{\frac{m_1 + \cdots + m_r}{r}}),
$$
completing the proof. $\Box$

Denote $K = K_{m, \underbrace{s, \ldots ,s}_{r}}$, where $m \geq 2$ and
$s \geq 2$. Every induced subgraph of $K$ has a vertex of degree at
most $rs$, and therefore by corollary~\ref{c44}
$ch_k(K) \leq k(rs+1)$ for all $k \geq 1$. Note that this upper bound 
for $ch_k(K)$ does not depend of $m$, which means that a good lower bound 
for $ch_k(K_{m_1, \ldots ,m_r})$ has a more complicated form than the 
upper bound given in theorem~\ref{t31}.

\noindent
{\bf Proof of Corollary~\ref{c31}}\,
Let $G=(V,E)$ be a graph and $k \geq 1$. Denote $r=\chi (G)$, and let
$V=V_1 \cup \ldots \cup V_r$ be a partition of the vertices, such that
each $V_i$ is a stable set. Denote $m_i=|V_i|$ for all 
$i$, $1 \leq i \leq r$. By theorem~\ref{t21}
$$
ch_k(G) \leq ch_k(K_{m_1+1, \ldots ,m_r+1}) \leq
948r(k+\log{\frac{m_1+ \cdots + m_r + r}{r}}) =
948\chi (G) (k+\log{(\frac{|V|}{\chi(G)}+1)}),
$$ 
as needed. $\Box$

\noindent
{\bf Proof of Corollary~\ref{c32}}\,
As proved by Bollob\'as in~\cite{Bo1}, for a fixed probability $p$,
$0 < p < 1$, almost surely (i.e., with probability that tends to $1$ as 
$n$ tends to infinity), the random graph $G_{n,p}$ has chromatic number 
$$
(\frac{1}{2}+o(1)) \log{(1/(1-p))}\frac{n}{\log n }.
$$
By corollary~\ref{c31}, for every $\epsilon > 0$ almost surely
$$
ch_k(G_{n,p}) \leq 948(\frac{1}{2}+\epsilon)\log{(1/(1-p))}\frac{n}{\log n}
(k+\log{(\frac{3\log n}{\log{(1/(1-p))}}+1)}).
$$
The result follows since $k$ and $p$ are constants. $\Box$

\noindent
Note that in the proof of the last corollary we have not used any knowledge 
concerning independent sets of $G_{n,p}$, as was done in~\cite{A2} for the
proof of the special case.

\section{Choice numbers and orientations}

Let $D=(V,E)$ be a digraph. We denote the set of out-neighbors of $v$ 
in $D$ by $N^+_D(v)$. A set of vertices $K \subseteq V$ is called a
{\em kernel} of $D$ if $K$ is an independent set and  
$N^+_D(v) \cap K \neq \emptyset$ for every vertex $v \not\in K$. 
Richardson's theorem (see, e.g.,~\cite{Be}) states that any digraph with
no odd directed cycle has a kernel.

\noindent
{\bf Proof of Theorem~\ref{t41}}\,
Let $D=(V,E)$ be a digraph which contains no odd directed (simple) cycle 
and $k \geq 1$. For each $v \in V$, let $S(v)$ be a set of size
$k(d^+_D(v)+1)$. We claim that the following algorithm finds subsets 
$C(v) \subseteq S(v)$, where $|C(v)|=k$ for all $v \in V$, and
$C(u) \cap C(v) = \emptyset$ for every two adjacent vertices $u,v \in V$.
\begin{enumerate}
\item
\label{s1}
$S \leftarrow \cup_{v \in V} S(v)$, $W \leftarrow V$ and for
every $v \in V$, $C(v) \leftarrow \emptyset$.
\item
\label{s2}
Choose a color $c \in S \cap \cup_{v \in W} S(v)$ and put
$S \leftarrow S - \{c\}$. 
\item
\label{s3}
Let $K$ be a kernel of the induced subgraph of $D$ on the vertex set
$\{ v \in W : c \in  S(v) \}$.
\item
\label{s4}
$C(v) \leftarrow C(v) \cup \{c\}$ for all $v \in K$.
\item
\label{s5}
$W \leftarrow W - \{ v \in K : |C(v)|=k \}$.
\item
\label{s6}
If $W = \emptyset$, stop. If not, go to step~\ref{s2}. 
\end{enumerate}

During the algorithm, $W$ is equal to $\{v \in V : |C(v)|<k \}$, and $S$ is
the set of remaining colors. We first prove that in step~\ref{s2},
$S \cap \cup_{v \in W} S(v) \neq \emptyset$. When the algorithm reaches
step~\ref{s2}, it is obvious that $W \neq \emptyset$. Suppose that $w \in W$ in 
this step, and therefore $|C(w)|<k$. It follows easily from the definition of
a kernel that every color from $S(w)$, which has been previously chosen in
step~\ref{s2}, belongs either to $C(w)$ or to $\cup_{v \in N^+_D(w)} C(v)$.
Since 
$$
|C(w)| + |\bigcup_{v \in N^+_D(w)} C(v)| < k + k \cdot d^+_D(v) = |S(w)|,
$$
not all the colors of $S(w)$ have been used. This means that 
$S \cap S(w) \neq \emptyset$, as needed. It follows easily that the
algorithm always terminates.

Upon termination of the algorithm, $|C(v)|=k$ for all $v \in V$. In 
step~\ref{s4} the same color is assigned to the vertices of a kernel which is 
an independent set, and therefore $C(u) \cap C(v) = \emptyset$ for every two 
adjacent vertices $u,v \in V$. This proves the correctness of the algorithm.

In step~\ref{s4}, the operation $C(v) \leftarrow C(v) \cup \{c\}$ is performed 
for at least one vertex. Upon termination $|\cup_{v \in V} C(v)| \leq k|V|$, 
which means that the algorithm performs at most $k|V|$ iterations. There is
a polynomial time algorithm for finding a kernel in a digraph with no odd 
directed cycle. Thus, the algorithm is of polynomial time complexity 
in $|V|$ and $k$, completing the proof. $\Box$

\noindent
{\bf Proof of Corollary~\ref{c41}}\,
This is an immediate consequence of theorem~\ref{t41}, since
$k(d^+_D(v)+1) \leq k(d+1)$ for every $v \in V$. $\Box$

\noindent
{\bf Proof of Corollary~\ref{c42}}\,
The result follows from~\ref{c41} by taking the cyclic orientation of
the even cycle. $\Box$

The proof of corollary~\ref{c43} is similar to the proof of the special
case which appears in~\cite{ERT}. A graph $G=(V,E)$ is
{\em $k$-degree-choosable} if for every family of sets $\{S(v): v \in V \}$,
where $|S(v)|=k d(v)$ for all $v \in V$, there are subsets 
$C(v) \subseteq S(v)$, where $|C(v)|=k$ for all $v \in V$, and 
$C(u) \cap C(v)= \emptyset$ for every two adjacent vertices $u,v \in V$.  
\begin{lemma}
\label{l41}
If a graph $G=(V,E)$ is connected, and $G$ has a connected induced subgraph 
$H=(V',E')$ which is $k$-degree-choosable, then $G$ is $k$-degree-choosable.
\end{lemma}
{\bf Proof}\,
For each $v \in V$, let $S(v)$ be a set of size $k d(v)$. The proof is 
by induction on $|V|$. In case $|V|=|V'|$ there is nothing to prove. Assuming
that $|V|>|V'|$, let $v$ be a vertex of $G$ which is at maximal distance   
from $H$. This guarantees that $G - v$ is connected. Choose any subset
$C(v) \subseteq S(v)$ such that $|C(v)|=k$, and remove the colors of
$C(v)$ from all the vertices adjacent to $v$. The choice can be completed
by applying the induction hypothesis on $G - v$. $\Box$
\begin{lemma}
\label{l42}
If $c \geq 2$, then $\Theta_{a,b,c}$ is $k$-degree-choosable for every
$k \geq 1$.
\end{lemma}
{\bf Proof}\,
Suppose that $\Theta_{a,b,c}$ has vertex set 
$V=\{u,v,x_1,\ldots,x_{a-1},y_1,\ldots,y_{b-1},z_1,\ldots,z_{c-1}\}$ and
contains the three paths 
$u-x_1-\cdots-x_{a-1}-v$, $u-y_1-\cdots-y_{b-1}-v$, and
$u-z_1-\cdots-z_{c-1}-v$. 
For each $w \in V$, let $S(w)$ be a set of size $kd(w)$. For the vertex
$u$ we choose a subset $C(u) \subseteq S(u) - S(z_1)$ of size $k$. 
For each node according to the sequence 
$x_1,\ldots,x_{a-1},y_1,\ldots,y_{b-1},v,z_{c-1},\ldots,z_1$ we choose  
a subset of $k$ colors that were not chosen in adjacent earlier nodes.
$\Box$

\noindent
For the proof of corollary~\ref{c43}, we shall need the following lemma
which appears in~\cite{ERT}.
\begin{lemma}
\label{l43}
If there is no node which disconnects $G$, then $G$ is an odd cycle, or   
$G=K_n$, or $G$ contains, as a node induced subgraph, an even cycle without
chord or with only one chord.
\end{lemma}     

\noindent
{\bf Proof of Corollary~\ref{c43}}\,
Suppose that a connected graph $G$ is not $K_n$, and not an odd cycle.
If $G$ is not a regular graph, then every induced subgraph of $G$ has 
a vertex of degree at most $\Delta(G)-1$, and by corollary~\ref{c46}
$ch_k(G) \leq k \Delta(G)$ for all $k \geq 1$. 
If $G$ is a regular graph, then there is  
a part of $G$ not disconnected by a node, which is neither an odd cycle
nor a complete graph. It follows from lemma~\ref{l43} that $G$ contains,
as a node induced subgraph, an even cycle or a particular kind of 
$\Theta_{a,b,c}$ graph. We know from corollary~\ref{c42} and lemma~\ref{l42}
that both an even cycle and $\Theta_{a,b,c}$ are $k$-degree-choosable for
every $k \geq 1$. The result follows from lemma~\ref{l41}. $\Box$

\noindent
{\bf Proof of Corollary~\ref{c44}}\,
It is proved in~\cite{AT} that a graph $G=(V,E)$ has an orientation $D$ in
which every outdegree is at most $d$ if and only if $M(G) \leq d$. Therefore,
there is an orientation $D$ of $G$ in which the maximum outdegree is at
most $\lceil M(G) \rceil$. Since $D$ contains no odd directed cycles, the
result follows from corollary~\ref{c41}. $\Box$

\noindent
{\bf Proof of Corollary~\ref{c45}}\,
$M(G) \leq 2$, since any bipartite (simple) graph on $r$ vertices contains
at most $2r-2$ edges. The result follows from corollary~\ref{c44}. $\Box$

\noindent
{\bf Proof of Corollary~\ref{c46}}\,
We claim that if every induced subgraph of a graph $G=(V,E)$ has a vertex of
degree at most $d$, then $G$ has an acyclic orientation in which the 
maximum outdegree is $d$. The proof is by induction on $|V|$. If $|V|=1$,
the result is trivial. If $|V|>1$, let $v$ be a vertex of $G$ with degree
at most $d$. By the induction hypothesis, $G - v$ has an acyclic orientation
in which the maximum outdegree is $d$. We complete this orientation of 
$G - v$ by orienting every edge incident to $v$ from $v$ to its appropriate
neighbor and obtain the desired orientation of $G$, as claimed. The result 
follows from corollary~\ref{c41}. $\Box$

An undirected graph $G$ is called {\em triangulated} if $G$ does not
contain an induced subgraph isomorphic to $C_n$ for $n \geq 4$. Being 
triangulated is a hereditary property inherited by all the induced 
subgraphs of $G$. A vertex $v$ of $G$ is called {\em simplicial} if its
adjacency set $Adj(v)$ induces a complete subgraph of $G$. It is proved 
in~\cite{G} that every triangulated graph has a simplicial vertex.

\noindent
{\bf Proof of Corollary~\ref{c47}}\,
Suppose that $G$ is a triangulated graph, and let $H$ be an induced subgraph
of $G$. Since $H$ is triangulated, it has a simplicial vertex $v$.
The set of vertices $\{v\} \cup Adj_H(v)$ induces a complete subgraph of $H$,
and therefore $v$ has degree at most $\omega(G)-1$ in $H$. It follows
from corollary~\ref{c44} that $ch_k(G) \leq k\omega(G)$ for
every $k \geq 1$.
For every graph $G$ and $k \geq 1$, $ch_k(G) \geq k\omega(G)$ and hence
$ch_k(G)=k\omega(G)$ for every $k \geq 1$. Since $G$ is triangulated, it
is also perfect, which means that $\chi(G)=\omega(G)$, as needed. $\Box$

\noindent
{\bf Proof of Corollary~\ref{c48}}\,
It is easy to see that $L(G)$ is triangulated if and only if $G$ contains
no $C_n$ for every $n \geq 4$. The result follows from
corollary~\ref{c47}. $\Box$

The validity of the list-chromatic conjecture for graphs of class $2$ with
maximum degree $3$ (and in particular for snarks) follows easily from
corollary~\ref{c43}. Suppose that $G$ is a graph of class $2$ with 
$\Delta(G)=3$. Let $C$ be a connected component of $L(G)$. If $C$ is not 
a complete graph, and not an odd cycle, then 
$ch(C) \leq \Delta(C) \leq \Delta(L(G)) \leq 4$.
If $C$ is a complete graph or an odd cycle, then it is easy to see
that $\Delta(C) \leq 2$, and therefore by corollary~\ref{c44}
$ch(C) \leq \Delta(C)+1 \leq 3$. It follows that $ch(L(G)) \leq 4$.
Since $G$ is a graph of class $2$,
$ch(L(G)) \geq \chi(L(G)) = \Delta(G)+1 = 4$, and hence,
$ch(L(G))= \chi(L(G)) = 4$.

\section{Properties of $(2k:k)$-choosable graphs}

Let $A$ and $B$ be sets of size $4$. We denote
$p(A,B)=\{(C,D):C \subseteq A, D \subseteq B, |C|=|D|=2\}$. Suppose
that $S \subseteq p(A_1,B_1)$ and that $T \subseteq p(A_2,B_2)$.
We say that $S$ and $T$ are isomorphic if there exist two bijections
$f:A_1 \mapsto A_2$ and $g:B_1 \mapsto B_2$ so that $(C,D) \in S$
iff $(f(C),g(D)) \in T$ for every $C \subseteq A$ and $D \subseteq B$,
where $|C|=|D|=2$.

Let $A$ and $B$ be sets of size $4$, and suppose that $S \subseteq p(A,B)$.
Suppose that $H_1, \ldots ,H_6$ are all the subsets of $A$ of size $2$.
For each $i$, $1 \leq i \leq 6$, we denote $c(H_i)=\{G: (H_i,G) \in S \}$
and $d_i=|c(H_i)|$. The sequence $(d_1, \ldots ,d_6)$ is called the degree
sequence of $S$. We say that $S$ is special if it has the
following properties:
\begin{enumerate}
\item
Its degree sequence is $(6,5,5,3,3,1)$.
\item
If $H$ and $G$ are the two subsets of $A$ for which
$|c(H)|=|c(G)|=3$, then $|H \cap G|=1$.
Denote $H=\{1,2\}$, $G=\{1,3\}$, and $A=\{1,2,3,4\}$.
\item
$c(H)=c(G)$.
\item
$c(H)$ has either the form
$\{ \{5,6\},\{5,7\},\{5,8\} \}$ or the form $\{ \{5,6\},\{5,7\},\{6,7\} \}$.
\item
Either $|c(\{2,3\})|=1$ and $|c(\{1,4\})|=6$, or
$|c(\{2,3\})|=6$ and $|c(\{1,4\})|=1$.
\end{enumerate}
We say that $S$ has property $P_1$ iff $comp(H)$ has the form
$\{ \{5,6\},\{5,7\},\{5,8\} \}$ and that it has property $P_2$ iff
$|comp(\{2,3\})|=1$.

Suppose that $K_{2,2}$ has vertex set $V=X \cup Y$, where
$X=\{x_1,x_2\}$, $Y=\{y_1,y_2\}$, and it has exactly
the edges $\{x_i,y_j\}$. For each $v \in V$, let $S(v)$ be
a set of size $4$. By $C(v)$ we denote a subset of $S(v)$ of size $2$.
We say that $C(x_1)$ and $C(x_2)$ are compatible
if there exist two subsets $C(y_1)$ and $C(y_2)$,
so that $C(u) \cap C(v) = \emptyset$ for every two adjacent
vertices $u,v \in V$. A subset $C(x_1) \subseteq S(x_1)$ is called
bad if $C(x_1)$ is not compatible with any $C(x_2)$. An analogous
definition is made for $C(x_2)$. We say that a family of sets
$\{S(v):v \in V\}$ is defected if there exist two bad subsets
$C(x_1)$ and $C(x_2)$. We denote by $incomp(x_1,x_2)$ the set of
incompatible pairs $(C(x_1),C(x_2))$.
\begin{lemma}
\label{l51}
If the family of sets $\{S(v):v \in V\}$ is defected and $C(x_1)$ is bad,
then both $S(y_1)$ and $S(y_2)$ intersect $C(x_1)$ and at least one them
contains $C(x_1)$.
\end{lemma}
{\bf Proof}\,
Suppose that neither $S(y_1)$ nor $S(y_2)$ contain $C(x_1)$. Remove the
colors of $C(x_1)$ from $S(y_1)$ and $S(y_2)$. Now both $S(y_1)$ and $S(y_2)$
have size at least $3$. We can assume the worst case, in which
both $S(y_1)$ and $S(y_2)$ are subsets of $S(x_2)$, and therefore
$|S(y_1) \cap S(y_2)| \geq 2$. Let $C$ be a subset of $S(y_1) \cap S(y_2)$
of size $2$. Choose a subset $C(x_2) \subseteq S(x_2)-C$. We have that
$C(x_1)$ and $C(x_2)$ are compatible in contrast to the fact that $C(x_1)$
is bad. This proves that at least one of $S(y_1)$ and $S(y_2)$ contains
$C(x_1)$.

Suppose that $S(y_1) \cap C(x_1) = \emptyset$. Choose a subset
$C(y_2) \subseteq S(y_2)-C(x_1)$ and a subset
$C(x_2) \subseteq S(x_2)-C(y_2)$. We have that $C(x_1)$ and $C(x_2)$ are
compatible in contrast to the fact that $C(x_1)$ is bad.
This proves that both $S(y_1)$ and $S(y_2)$ intersect $C(x_1)$. $\Box$
\begin{lemma}
\label{l52}
If the family of sets $\{S(v):v \in V\}$ is defected, then both $S(x_1)$
and $S(x_2)$ contain exactly one bad subset. Furthermore, at least one of
the following is valid:
\begin{enumerate}
\item
The set $incomp(x_1,x_2)$ is special and has properties $P_1$ and $P_2$.
\item
$incomp(x_1,x_2)$ has degree sequence $(6,5,5,3,2,2)$.
\item
$|incomp(x_1,x_2)|=21$.
\end{enumerate}
\end{lemma}
{\bf Proof}\,
The set $S(x_1)$ contains a bad subset, which we denote by $C(x_1)=\{1,2\}$.
Without loss of generality, we can assume by lemma~\ref{l51} that
$C(x_1) \subseteq S(y_1)$ and that $S(y_2)$ intersects $C(x_1)$.
Denote $S(y_1)=\{1,2,3,4\}$. Since $C(x_1)$ is bad, we must have
that $|(S(y_1) \cap S(y_2))-C(x_1)|<2$.

\noindent
{\bf Case 1:}\, $C(x_1) \subseteq S(y_2)$ and $|S(y_1) \cap S(y_2)|=3$.

\noindent
Denote $S(y_2)=\{1,2,3,5\}$. Since $C(x_1)$ is bad, surely
$\{3,4,5\} \subseteq S(x_2)$. The set $S(x_2)$ contains a bad subset, which
we denote by $C(x_2)$, and therefore $\{1,2\} \cap S(x_2) \neq \emptyset$.
We can assume, without loss of generality, that $S(x_2)=\{1,3,4,5\}$.
Since $C(x_2)$ is bad and $|S(y_1) \cap S(y_2)|=3$, we must have that
$C(x_2) \subseteq S(y_1) \cap S(y_2)$. Hence, $C(x_2)=\{1,3\}$ and
$S(x_1)=\{1,2,4,5\}$. We have that
$$
S(x_1)=\{1,2,4,5\},S(x_2)=\{1,3,4,5\},S(y_1)=\{1,2,3,4\},S(y_2)=\{1,2,3,5\}.
$$
The set $incomp(x_1,x_2)$ is special and has properties $P_1$ and $P_2$.

\noindent
{\bf Case 2:}\, $C(x_1) \subseteq S(y_2)$ and $|S(y_1) \cap S(y_2)|=2$.

\noindent
Denote $S(y_2)=\{1,2,5,6\}$. Since $C(x_1)$ is bad, surely
$|S(x_2) \cap \{3,4,5,6\}| \geq 3$. Suppose without loss of generality
that $\{3,4,5\} \subseteq S(x_2)$. The set $S(x_2)$ contains a bad subset,
which we denote by $C(x_2)$, and therefore
$\{1,2\} \cap S(x_2) \neq \emptyset$. We can assume, without loss of
generality, that $S(x_2)=\{1,3,4,5\}$. Since $C(x_2)$ is bad and
$|S(y_1) \cap S(y_2)|=2$, we must have that
$C(x_2) \cap \{1,2\} \neq \emptyset$, and therefore $1 \in C(x_2)$.
We can assume, without loss of generality, that $C(x_2)=\{1,3\}$.
Since $C(x_2)$ is bad, we must have that $4 \in S(x_1)$ and
$S(x_1) \cap \{5,6\} \neq \emptyset$. Suppose without loss of generality
that $S(x_1)=\{1,2,4,5\}$. This is a contradiction to the fact that
$C(x_2)$ is bad.

\noindent
{\bf Case 3:}\, $|C(x_1) \cap S(y_2)|=1$ and $|S(y_1) \cap S(y_2)|=2$.

\noindent
We can assume, without loss of generality, that $1 \in S(y_2)$.
Denote $S(y_2)=\{1,3,5,6\}$. Since $C(x_1)$ is bad, surely
$S(x_2)=\{3,4,5,6\}$. The set $S(x_2)$ contains a bad subset, which we
denote by $C(x_2)$. Since $C(x_2)$ is bad and $|S(y_1) \cap S(y_2)|=2$,
we must have that $C(x_2) \cap \{1,3\} \neq \emptyset$, and therefore
$3 \in C(x_2)$. If $C(x_2)=\{3,4\}$, then we must have that
$S(x_1)=\{1,2,5,6\}$, so
$$
S(x_1)=\{1,2,5,6\},S(x_2)=\{3,4,5,6\},S(y_1)=\{1,2,3,4\},S(y_2)=\{1,3,5,6\}.
$$
The set $incomp(x_1,x_2)$ has degree sequence $(6,5,5,3,2,2)$.
Otherwise, suppose without loss of generality that $C(x_2)=\{3,5\}$. We
must have that $S(x_1)=\{1,2,4,6\}$, so
$$
S(x_1)=\{1,2,4,6\},S(x_2)=\{3,4,5,6\},S(y_1)=\{1,2,3,4\},S(y_2)=\{1,3,5,6\}.
$$
In this case $|incomp(x_1,x_2)|=21$.

\noindent
{\bf Case 4:}\, $|C(x_1) \cap S(y_2)|=1$ and $|S(y_1) \cap S(y_2)|=1$.

\noindent
We can assume, without loss of generality, that $1 \in S(y_2)$.
Denote $S(y_2)=\{1,5,6,7\}$. Since $C(x_1)$ is bad, we must have that
$S(x_2) \cap \{3,4\} \neq \emptyset$ and $|S(x_2) \cap \{5,6,7\}| \geq 2$.
Suppose without loss of generality that $\{3,5,6\} \subseteq S(x_2)$.
Since $C(x_1)$ is bad, we must have that either $S(x_2)=\{4,3,5,6\}$ or
$S(x_2)=\{7,3,5,6\}$. It is easy to see that in both cases we have
a contradiction to the fact that $S(x_2)$ contains a bad subset. $\Box$

For every $i$, $1 \leq i \leq m$, let $A_i$ be a sequence of $4$ distinct
elements. The sequence $A_1, \ldots ,A_m$ is called valid if whenever
$c \in A_i \cap A_{i+1}$, then $c$ appears in the same position in both
$A_i$ and $A_{i+1}$. A valid sequence $A_1, \ldots , A_m$ is called
legal if whenever $c \in A_{i+1}-A_i$, then $c \not\in A_j$ for every $j$, 
$1 \leq j \leq i$. By a subsequence of $A_1, \ldots ,A_m$ we mean a sequence 
of the form $A_i,A_{i+1}, \ldots ,A_j$, where $1 \leq i \leq j \leq m$.

Let $A_1, \ldots ,A_m$ be a valid sequence. The pair $(A_i,A_{i+1})$ contains
a change in the $k$th position if the elements which appear in the $k$th
position of $A_i$ and $A_{i+1}$ are different. The sequence $A_1, \ldots ,A_m$
contains a change in the $k$th position if there exists a pair $(A_i,A_{i+1})$
which contains a change in the $k$th position.

Let $A_1, \ldots ,A_m$ be a sequence. By $C_i$ we denote a subset of
$A_i$ of size $2$. We say that $C_1$ and $C_m$ are compatible if there
exist subsets $\{C_k: 1 < k < m\}$ so that $C_p \cap C_{p+1} = \emptyset$
for every $p$, $1 \leq p < m$. A subset $C_1$ is called bad if $C_1$ is not
compatible with any $C_m$. A subset $C_1$ is called good if $C_1$ is
compatible with every $C_m$. We denote by $comp(C_1;A_1, \ldots ,A_m)$ the
set which consists of all the subsets $C_m$ which are compatible with $C_1$,
and by $comp(A_1, \ldots ,A_m)$ the set of all the compatible pairs
$(C_1,C_m)$. By $good(A_1, \ldots ,A_m)$ we denote the set which consists of
all the good subsets that $A_1$ contains.
\begin{lemma}
\label{l53}
If the valid sequence $D_1, \ldots ,D_r$ contains a change in at least $3$
positions and there is no $i$, $1 < i < r-1$, for which $D_i=D_{i+1}$, then
it contains a subsequence $A_1, \ldots ,A_m$, so that the sequence $A_1$
contains at least one good subset.
Furthermore, the sequence $A_1, \ldots ,A_m$ has at least one of the
following properties:
\begin{enumerate}
\item
$|good(A_1, \ldots ,A_m)| \geq 3$.
\item
$|comp(A_1, \ldots ,A_m)| > 23$.
\item
The set $comp(A_1, \ldots ,A_m)$ is special. If $m$ is odd, then
$comp(A_1, \ldots ,A_m)$ has exactly one of the properties $P_1$ and $P_2$.
If $m$ is even then $comp(A_1, \ldots ,A_m)$ has either both or none of the
properties $P_1$ and $P_2$.
\end{enumerate}
\end{lemma}
{\bf Proof}\,
We consider the following cases.

\noindent
{\bf Case 1:}\, For some $i$, $|D_i \cap D_{i+1}| \leq 1$.

\noindent
In this case $|good(D_i,D_{i+1})| \geq 3$.

\noindent
{\bf Case 2:}\, For every $j$, $|D_j \cap D_{j+1}| \leq 2$, and for
some $i$, $|D_i \cap D_{i+1}|=2$.

\noindent
Assume without loss of generality that the pair $(D_i,D_{i+1})$ contains
a change in the first and second positions. At least one of the pairs
$(D_{i-1},D_i)$ and $(D_i,D_{i+1})$ contains a change in some position.
Suppose that the pair $(D_{i-1},D_i)$ contains a change in some position.
The proof in case the pair $(D_i,D_{i+1})$ contains a change in some position
is similar.
If the pair $(D_{i-1},D_i)$ contains a change in at least one of the first
and second positions, then surely $|good(D_{i-1},D_i,D_{i+1})| \geq 3$.
If the only position in which the pair $(D_{i-1},D_i)$ contains a change
is either the third or the fourth position, then $comp(D_{i-1},D_i,D_{i+1})$
is special, has property $P_2$, and does not have property $P_1$.
If the pair $(D_{i-1},D_i)$ contains a change in the third and fourth
positions, then $|comp(D_{i-1},D_i,D_{i+1})|=27$.

\noindent
{\bf Case 3:}\, For every $j$, $|D_j \cap D_{j+1}| \leq 1$.

\noindent
Let $B_1, \ldots B_k$ be a subsequence of $D_1, \ldots ,D_r$
which contains a change in at least $3$ positions, but no proper subsequence
of $B_1, \ldots ,B_k$ has this property. This implies that the
three pairs $(B_1,B_2)$, $(B_2,B_3)$ and $(B_{k-1},B_k)$ contain a change
in three different positions. We can assume, without loss of generality,
that the three pairs contain a change in the first, second and third
positions respectively. Suppose that $2 \leq i \leq k-2$, and consider
the pair $(B_i,B_{i+1})$. If this pair contains a change in the first position,
then the sequence $B_2, \ldots ,B_m$ contains a change in at least
$3$ positions. If this pair contains a change in the third or fourth position,
then the sequence $B_1, \ldots ,B_{i+1}$ contains a change in at least
$3$ positions. Hence, the pair $(B_i,B_{i+1})$ contains
a change in the second position.
If $k=4$ then the set $comp(B_1, \ldots ,B_4)$ is special and does not
have neither property $P_1$ nor property $P_2$.
If $k>4$ then $|good(B_1, \ldots ,B_4)|=3$. $\Box$
\begin{lemma}
\label{l54}
If the set $comp(A_1, \ldots ,A_m)$ is special, then both the set
$comp(A_1,A_1, \ldots ,A_m)$ and the set $comp(A_1, \ldots ,A_m,A_m)$
are special. 
The set $comp(A_1,A_1, \ldots ,A_m)$ has property $P_1$ iff the set 
$comp(A_1, \ldots ,A_m)$ has property $P_1$.
The set $comp(A_1,A_1, \ldots ,A_m)$ has property $P_2$ iff the set
$comp(A_1, \ldots ,A_m)$ does not have property $P_2$.
The set $comp(A_1, \ldots ,A_m,A_m)$ has property $P_1$ iff the set
$comp(A_1, \ldots ,A_m)$ does not have property $P_1$.
The set $comp(A_1, \ldots ,A_m,A_m)$ has property $P_2$ iff the set
$comp(A_1, \ldots ,A_m)$ has property $P_2$.
\end{lemma}
\begin{lemma}
\label{l55}
If $A_1,A_2,A_2,A_3$ is a legal sequence, then
$$
comp(A_1,A_2,A_2,A_3)=comp(A_1,A_3).
$$
\end{lemma}
{\bf Proof}\,
Let $k_1, \ldots k_n$ be all the positions in which $A_1,A_2,A_2,A_3$
does not contain a change. It is easy to verify that
$(C,D) \in comp(A_1,A_3)$ iff there is no $i$ for which $C$ contains
the $k_i$th element of $A_1$ and $D$ contains the $k_i$ element of $A_3$.
The same property holds also for $comp(A_1,A_2,A_2,A_3)$. $\Box$
\begin{lemma}
\label{l56}
If $A_i, \ldots ,A_j$ is a subsequence of $A_1, \ldots ,A_m$, then
$$
|comp(A_1, \ldots ,A_m)| \geq |comp(A_i, \ldots ,A_j)|.
$$
\end{lemma}
{\bf Proof}\,
By induction on $m$. If $m=j-i+1$, there is nothing to prove. Suppose that
$m>j-i+1$. Assume that $i>1$. The proof in case $j<m$ is similar. Hence,
$$
|comp(A_1, \ldots ,A_m)| \geq |comp(A_2,A_2, \ldots ,A_m)| =
|comp(A_2, \ldots ,A_m)| \geq |comp(A_i, \ldots ,A_j)|,
$$
where the last inequality follows from the induction hypothesis. $\Box$
\begin{lemma}
\label{l57}
If $A_i, \ldots ,A_j$ is a subsequence of $A_1, \ldots ,A_m$, then
$$
|good(A_1, \ldots ,A_m)| \geq |good(A_i, \ldots ,A_j)|.
$$
\end{lemma}
{\bf Proof}\,
Similar to the proof of lemma~\ref{l56}. $\Box$
\begin{lemma}
\label{l58}
Suppose that $i \geq 0$, and denote by $F$ the sequence
$A_{i+1}, \ldots ,A_m$ together with an additional $A_{i+1}$ as the first
element of the sequence in case $i \equiv 1 \pmod{2}$. If
$A_{i+1}, \ldots ,A_m$ is a subsequence of $A_1, \ldots ,A_m$ and
$|comp(A_{i+1}, \ldots ,A_m)|=|comp(A_1, \ldots ,A_m)|$, then
$comp(A_1 \ldots ,A_m)$ is isomorphic to $comp(F)$.
\end{lemma}
{\bf Proof}\,
We can assume that $A_1, \ldots , A_m$ is a valid sequence.
Suppose that $i \equiv 1 \pmod{2}$. The proof in case $i \equiv 0 \pmod{2}$
is similar. Suppose that $C_1 \subseteq A_1$. Denote by $T$ the
subset of $A_{i+1}$ that appears in the two positions in which $C_1$ does
not appear in $A_1$. Since $A_1, \ldots ,A_{i+1}$ is a valid sequence,
we have that $C_1$ is compatible with $T$. Hence,
$$
V=comp(C_1;A_1, \ldots ,A_m) \supseteq comp(T;A_{i+1}, \ldots ,A_m)=W.
$$
Since $|comp(A_{i+1}, \ldots ,A_m)|=|comp(A_1, \ldots ,A_m)|$, we must have
that $V=W$. It is easy to see now that $comp(A_1, \ldots ,A_m)$ is
isomorphic to $comp(A_{i+1},A_{i+1}, \ldots ,A_m)$. $\Box$
\begin{lemma}
\label{l59}
Suppose that $i,j \geq 0$. Denote by $F$ the sequence
$A_{i+1}, \ldots ,A_{m-j}$ together with an additional $A_{i+1}$ as the first
element of the sequence in case $i \equiv 1 \pmod{2}$ and an additional
$A_{m-j}$ as the last element of the sequence in case $j \equiv 1 \pmod{2}$.
If $A_{i+1}, \ldots ,A_{m-j}$ is a subsequence of $A_1, \ldots ,A_m$ and
$|comp(A_{i+1}, \ldots ,A_{m-j})|=|comp(A_1, \ldots ,A_m)|$, then
$comp(A_{i+1} \ldots ,A_{m-j})$ is isomorphic to $comp(F)$.
\end{lemma}
{\bf Proof}\,
Apply lemma~\ref{l58} twice. $\Box$
\begin{lemma}
\label{l510}
Suppose that $r$ is odd and that $r \geq 3$. If the valid sequence
$D_1, \ldots ,D_r$ contains a change in at least $3$ positions,
then the sequence $D_1$ contains at least one good subset.
Furthermore, at least one of the following is valid:
\begin{enumerate}
\item
$|good(D_1, \ldots ,D_r)| \geq 3$.
\item
$|comp(D_1, \ldots ,D_r)| > 23$.
\item
The set $comp(D_1, \ldots ,D_m)$ is special and has exactly one of the
properties $P_1$ and $P_2$.
\end{enumerate}
\end{lemma}
{\bf Proof}\,
We can assume, without loss of generality, that $D_1, \ldots ,D_r$ is legal.
Due to lemma~\ref{l55}, we can assume that there is no $i$, $1 < i < r-1$,
for which $D_i=D_{i+1}$. It follows from lemma~\ref{l53} that the sequence
$D_1, \ldots ,D_r$ contains a subsequence $A_1, \ldots ,A_m$, so that
the sequence $A_1$ contains at least one good subset. It follows from
lemma~\ref{l57} that $|good(D_1, \ldots ,D_r)| \geq 1$.
According to lemma~\ref{l53}, we consider the following cases:

\noindent
{\bf Case 1:}\, $|good(A_1, \ldots ,A_m)| \geq 3$.

\noindent
It follows from lemma~\ref{l57} that $|good(D_1, \ldots ,D_r)| \geq 3$.

\noindent
{\bf Case 2:}\, $|comp(A_1, \ldots ,A_m)| \geq 27$.

\noindent
It follows from lemma~\ref{l56} that $|comp(D_1, \ldots ,D_r)| > 23$.

\noindent
{\bf Case 3:}\, The set $comp(A_1, \ldots ,A_m)$ is special.

\noindent
We know that if $m$ is odd, then $comp(A_1, \ldots ,A_m)$ has exactly one
of the properties $P_1$ and $P_2$. Furthermore, if $m$ is even then
$comp(A_1, \ldots ,A_m)$ has either both or none of the properties
$P_1$ and $P_2$. If $|comp(D_1, \ldots ,D_r)|>|comp(A_1, \ldots ,A_m)|$,
then $|comp(D_1, \ldots ,D_m)| > 23$. Suppose that
$|comp(D_1, \ldots ,D_r)|=|comp(A_1, \ldots ,A_m)|$. It follows from
lemma~\ref{l59} that $comp(D_1, \ldots ,D_r)$ is isomorphic to $comp(F)$ for
some sequence $F$. Since $r$ is odd and using lemma~\ref{l54}, it is easy to
see that $comp(D_1, \ldots ,D_r)$ is special and has exactly one of the
properties $P_1$ and $P_2$. $\Box$

\noindent
{\bf Proof of Theorem~\ref{t52}}\,
It is easy to see that a graph $G$ is $(4:2)$-choosable iff its core
is $(4:2)$-choosable. Due to theorem~\ref{t51}, we need to prove that
for every $m \geq 1$, $\Theta_{2,2,2m}$ is $(4:2)$-choosable. Suppose
that $m$ is odd and that $m \geq 3$. Assume that $\Theta_{2,2,m-1}$ has
vertex set $V=\{u,v,z_1, \ldots ,z_m\}$ and contains the three paths
$z_1 - z_2 - \cdots - z_m$, $z_1 - u - z_m$, and $z_1 - v - z_m$. For each
$w \in V$, let $S(w)$ be a set of size $4$. We denote $A_i=S(z_i)$ for every
$i$, $1 \leq i \leq m$. We can assume that $A_1, \ldots ,A_m$
is a valid sequence.

Suppose first that the sequence $A_1, \ldots ,A_m$ contains a change
in at most $2$ positions. This means that there is a set $C$ of size $2$
so that $C \subseteq A_i$ for every $i$, $1 \leq i \leq m$.
From $A_i$ when $i$ is odd, choose the subset $C$. Complete the choice
by choosing a subset of $S(w)-C$ for every other vertex $w$.

Suppose next that the sequence $A_1, \ldots ,A_m$ contains a change in
at least $3$ positions. The graph induced by the set of vertices
$\{z_1,z_m,u,v\}=W$ is isomorphic to $K_{2,2}$. Denote $x_1=z_1$,
$x_2=z_m$, $y_1=u$, and $y_2=v$. We use the same terminology as
before.

\noindent
{\bf Case 1:}\, $\{S(w):w \in W\}$ is not defected.

\noindent
Suppose without loss of generality that $S(z_1)$ contains no bad subset.
If follows from lemma~\ref{l510} that $|good(A_1, \ldots ,A_m)| \geq 1$, 
and therefore a choice is possible.

\noindent
{\bf Case 2:}\, $\{S(w):w \in W\}$ is defected.

\noindent
According to lemma~\ref{l510}, we consider the following cases:

\noindent
{\bf Case 2a:}\, $|good(A_1, \ldots ,A_m)| \geq 3$.

\noindent
It follows from lemma~\ref{l52} that $S(z_1)$ contains exactly one bad subset,
and therefore a choice is possible.

\noindent
{\bf Case 2b:}\, $|comp(D_1, \ldots ,D_r)| > 23$.

\noindent
It follows from lemma~\ref{l52} that $|incomp(z_1,z_m)| \leq 23$, and
therefore a choice is possible.

\noindent
{\bf Case 2c:}\, The set $comp(D_1, \ldots ,D_m)$ is special.

\noindent
We know that $comp(D_1, \ldots ,D_m)$ has exactly one of the properties 
$P_1$ and $P_2$. It is easy to see from lemma~\ref{l52} that the set 
$incomp(z_1,z_m)$ does not contain the set $comp(D_1, \ldots ,D_m)$,
and therefore a choice is possible. $\Box$

\noindent
{\bf Proof of Theorem~\ref{t53}}\,
Suppose that $G=(V,E)$ is $(2mk:mk)$-choosable for $k$ odd. We prove that 
$G$ is $2m$-choosable as well. For each $v \in V$, let $S(v)$ be a set of
size $2m$. With every color $c$ we associate a set $F(c)$ of size $k$, such
that $F(c) \cap F(d) = \emptyset$ if $c \neq d$. For every $v \in V$,
we define $T(v)=\cup_{c \in S(v)} F(c)$. Since $G$ is $(2mk:mk)$-choosable, 
there are subsets $C(v) \subseteq T(v)$, where $|C(v)|=mk$ for 
all $v \in V$, and $C(u) \cap C(v) = \emptyset$ for every
two adjacent vertices $u,v \in V$.

Fix a vertex $v \in V$. Since $k$ is odd, there is a color $c \in S(v)$
for which $|C(v) \cap F(c)| > k/2$, so we define $f(v)=c$. 
In case $u$ and $v$ are adjacent vertices for which $c \in S(u) \cap S(v)$,
it is not possible that both $|C(u) \cap F(c)|$ and $|C(v) \cap F(c)|$ are
greater than $k/2$. This proves that $f$ is a proper vertex-coloring of
$G$ assigning to each vertex $v \in V$ a color in $S(v)$. $\Box$

\section{The complexity of graph choosability}

Let $G=(V,E)$ be a graph. We denote by $G'$ the graph obtained from $G$ by
adding a new vertex to $G$, and joining it to every vertex in $V$.  
Consider the following decision problem:

\vspace{3 mm}
\noindent
{\bf GRAPH $k$-COLORABILITY} \\
INSTANCE: A graph $G=(V,E)$. \\
QUESTION: Is $G$ $k$-colorable?

\vspace{3 mm}
\noindent
The standard technique to show a polynomial transformation from
GRAPH $k$-COLORABILITY to GRAPH $(k+1)$-COLORABILITY is to use the fact 
that $\chi(G') = \chi(G)+1$ for every graph $G$. 
However, it is not true that $ch(G') = ch(G)+1$ for every graph $G$.
To see that, we first prove that $K_{2,4}'$ is $3$-choosable.  

Suppose that $K_{2,4}'$ has vertex set $V=\{v,x_1,x_2,y_1,y_2,y_3,y_4\}$,
and contains exactly the edges $\{x_i,y_j\}$, $\{v,x_i\}$, and $\{v,y_j\}$.
For each $w \in V$, let $S(w)$ be a set of size $3$. 

\noindent
{\bf Case 1:}\, All the sets are the same.

\noindent
A choice can be made since $K_{2,4}'$ is 3-colorable.

\noindent
{\bf Case 2:}\, There is a set $S(x_i)$ which is not equal to $S(v)$.

\noindent
Without loss of generality, suppose that $S(v) \neq S(x_1)$.
For the node $v$, choose a color $c \in S(v)-S(x_1)$, and remove $c$ from
the sets of the other vertices. We can assume that every set $S(y_j)$
is of size $2$ now.

Suppose first that $S(x_1)$ and $S(x_2)$ are disjoint. The number of
different sets consisting of one color from each of the $S(x_i)$ is
at least 6, and therefore we can choose colors $c_i \in S(x_i)$, such 
that $\{c_1,c_2\}$ does not appear as a set of $S(y_j)$.  
We complete the choice by choosing for every vertex $y_j$ a color 
from $S(y_j)-\{c_1,c_2\}$. 
Suppose next that $c \in S(x_1) \cap S(x_2)$. For every vertex $x_i$ we 
choose $c$, and for every vertex $y_j$ we choose a color from $S(y_j)-\{c\}$.

\noindent
{\bf Case 3:}\, There is a set $S(y_j)$ which is not equal to $S(v)$.

\noindent
Without loss of generality, suppose that $S(v) \neq S(y_1)$.
For the node $v$, choose a color $c \in S(v)-S(y_1)$, and remove $c$ from 
the sets of the other vertices. 
Suppose first that $S(x_1)$ and $S(x_2)$ are disjoint. The number of
different sets consisting of one color from each of the $S(x_i)$ is 
at least 4, and since $|S(y_1)|=3$ we can choose colors $c_i \in S(x_i)$, 
such that $S(y_j)-\{c_1,c_2\} \neq \emptyset$ for every vertex $y_j$.
We can complete the choice as in case 2.
In case $S(x_1)$ and $S(x_2)$ are not disjoint, we proceed as in case 2. 

This completes the proof that $K_{2,4}'$ is 3-choosable. It follows from
theorem~\ref{t51} and corollary~\ref{c46} that $ch(K_{2,4})=3$, and
therefore $ch(K_{2,4}')=ch(K_{2,4})=3$. The following lemma exhibits
a construction which increases the choice number of a graph in exactly $1$.
\begin{lemma}
\label{l61}
Let $G=(V,E)$ be a graph. If $H$ is the disjoint union of $|V|$ copies of $G$,
then $ch(H')=ch(G)+1$.
\end{lemma}
{\bf Proof}\,
Let $H$ be the disjoint union of the graphs $\{G_i : 1 \leq i \leq |V|\}$, 
where each $G_i$ is a copy of $G$. 
Suppose that $H'$ is obtained from $H$ by joining the new vertex $v$ to all
the vertices of $H$.

We claim that if $G$ is $k$-choosable, then $H'$ is $(k+1)$-choosable.
For each $w \in V(H')$, let $S(w)$ be a set of size $k+1$. Choose a color
$c \in S(v)$, and remove $c$ from the sets of the other vertices. We can
complete the choice since $G$ is $k$-choosable.

We now prove that if $H'$ is $k$-choosable, then 
$G$ is $(k-1)$-choosable. It is easy to see that this is true when $G$ 
is a complete graph. If $G$ is not a complete graph, then by 
corollary~\ref{c43} $ch(G) < |V|$, and therefore $ch(H') \leq |V|$.
Hence, we can assume that $k \leq |V|$. 
For each $w \in V$, let $S(w)$ be a set of size 
$k-1$, such that $S(w) \cap \{1,2, \ldots , |V|\} = \emptyset$. 
For every $i$, $1 \leq i \leq |V|$, on the vertices of the graph $G_i$ we 
put the sets $S(w)$ together with the additional color $i$. The 
vertex $v$ is given the set $\{1,2, \ldots ,k\}$. 
Let $f$ be a proper vertex-coloring of $H'$ assigning to each vertex 
a color from its set. Denote $f(v)=i$, then $f$ restricted to $G_i$ is
a proper vertex-coloring of $G$ assigning to each vertex $w \in V$
a color in S(w). $\Box$  
\begin{lemma}
\label{l62}
{\bf BIPARTITE GRAPH $3$-CHOOSABILITY} is $\Pi_2^p$-complete.
\end{lemma}
{\bf Proof}\,
It is easy to see that {\bf BG $3$-CH} $\in \Pi_2^p$.
We transform {\bf BG $(2,3)$-CH} to {\bf BG $3$-CH}. 
Let $G=(V,E)$ and $f: V \mapsto \{2,3\}$ be an
instance of {\bf BG $(2,3)$-CH}.  
We shall construct a bipartite graph $W$ such that $W$ is $3$-choosable
if and only if $G$ is $f$-choosable.

Let $H$ be the disjoint union of the graphs $\{G_{i,j} : 1 \leq i,j \leq 3\}$,
where each $G_{i,j}$  is a copy of $G$. Let $(X,Y)$ be a bipartition of the
bipartite graph $H$. The graph $W$ is obtained from $H$ by adding two new
vertices $u$ and $v$, joining $u$ to every vertex $w \in X$ for which 
$f(w)=2$, and joining $v$ to every vertex $w \in Y$ for which $f(w)=2$.

Since $H$ is bipartite, $W$ is also a bipartite graph. It is easy to see that 
if $G$ is $f$-choosable, then $W$ is $3$-choosable. We now prove that
if $W$ is $3$-choosable, then $G$ is $f$-choosable. 
For every $w \in V$, let $S(w)$ be a set of size $f(w)$, such that 
$S(w) \cap \{1,2,3\} = \emptyset$.  
For every $i$ and $j$, $1 \leq i,j \leq 3$, on the vertices of the graph
$G_{i,j}$ we put the sets $S(w)$ with the vertices for which $f$ is equal
to $2$ receiving another color as follows:
to the vertices which belong to $X$ we add the color $i$, 
whereas to the vertices which belong to $Y$ we add the color $j$. 
The vertices $u$ and $v$ are both given the set $\{1,2,3\}$.
Let $f$ be a proper vertex-coloring of $H'$ assigning to each vertex
a color from its set. Denote $f(u)=i$ and $f(v)=j$, then $f$ restricted 
to $G_{i,j}$ is a proper vertex-coloring of $G$ assigning to each vertex
$w \in V$ a color in $S(w)$. $\Box$ 

\noindent
{\bf Proof of Theorem~\ref{t61}}\,
The proof is by induction on $k$. For $k=3$, the result follows from
lemma~\ref{l62}. Assuming that the result is true for $k$, $k \geq 3$, we
prove it is true for $k+1$. It is easy to see that 
{\bf BG $(k+1)$-CH} $\in \Pi_2^p$. We transform {\bf BG $k$-CH} to
{\bf BG $(k+1)$-CH}. Let $G=(V,E)$ be an instance of {\bf BG $k$-CH}.
We shall construct a bipartite graph $W$ such that $W$ is $(k+1)$-choosable
if and only if $G$ is $k$-choosable. 

Let $H$ be the disjoint union of the graphs 
$\{G_{i,j} : 1 \leq i,j \leq (k+1)^2\}$, where each $G_{i,j}$ is a copy 
of $G$. Let $(X,Y)$ be a bipartition of the bipartite graph $H$. 
The graph $W$ is obtained from $H$ by adding two new vertices 
$u$ and $v$, joining $u$ to every vertex of $X$, and joining $v$ to
every vertex of $Y$.

It is easy to see that if $G$ is $k$-choosable, then $W$ is $(k+1)$-choosable.
In a similar way to the proof of lemma~\ref{l62}, we can prove that if $W$ is
$(k+1)$-choosable, then $G$ is $k$-choosable. $\Box$

\section{The strong choice number}

Let $G=(V,E)$ be a graph, and let $V_1, \ldots ,V_r$ be pairwise disjoint
subsets of $V$. We denote by $[G,V_1, \ldots ,V_r]$ the graph obtained
from $G$ by adding to it the union of cliques induces by each  
$V_i$, $1 \leq i \leq r$.

Suppose that $G=(V,E)$ is a graph with maximum degree at most $1$. We claim
that $G$ is strongly $k$-choosable for every $k \geq 2$. To see that, 
let $V_1, \ldots ,V_r$ be pairwise disjoint subsets of $V$, each of size
at most $k$. The graph $[G,V_1, \ldots ,V_r]$ has maximum degree at   
most $k$, and therefore by corollary~\ref{c43} it is $k$-choosable.

\noindent
{\bf Proof of Theorem~\ref{t71}}\,
Let $G=(V,E)$ be a strongly $k$-colorable graph. Let $V_1, \ldots ,V_r$ be
pairwise disjoint subsets of $V$, each of size at most $k+1$. 
Without loss of generality, we can assume that $V_1, \ldots ,V_m$ are 
subsets of size exactly $k+1$, and $V_{m+1}, \ldots ,V_r$ are subsets
of size less than $k+1$. Let $H$ be the graph $[G,V_1, \ldots ,V_r]$. 
To complete the proof, it suffices to show that $H$ is $(k+1)$-colorable.  
For every $i$, $1 \leq i \leq m$, we define $W_i=V_i-\{c\}$ for an     
arbitrary element $c \in V_i$, whereas for every $j$, $m+1 \leq j \leq r$,
we define $W_i=V_i$. Since $[G,W_1, \ldots ,W_r]$ is $k$-colorable, there 
exists an independent set $S$ of $H$ which is composed of exactly one vertex
from each $V_i$, $1 \leq i \leq m$. For every $i$, $1 \leq i \leq m$, we
define $W_i=V_i-S$, whereas for every $j$, $m+1 \leq j \leq r$, we 
define $W_i=V_i$. Since $[G,W_1, \ldots ,W_r]$ is $k$-colorable, we can
obtain a proper $(k+1)$-vertex coloring of $H$ by using $k$ colors for
$V-S$ and another color for $S$. $\Box$

\begin{lemma}
\label{l71}
Suppose that $k,l \geq 1$. If ${\cal F}$ is a family of $k+l$ sets
of size $k+l$, then it is possible to partition ${\cal F}$ into a family 
${\cal F}_1$ of $k$ sets and a family ${\cal F}_2$ of $l$ sets, to 
choose for each set $S \in {\cal F}_1$ a subset $S' \subseteq S$ of size $k$,
and to choose for each set $T \in {\cal F}_2$ a subset $T' \subseteq T$ of
size $l$, so that $S' \cap T' = \emptyset$ for every $S \in {\cal F}_1$
and $T \in {\cal F}_2$.
\end{lemma}
{\bf Proof}\,
Suppose that ${\cal F} = \{C_1, \ldots , C_{k+l} \}$, and define 
$C = \cup_{i=1}^{k+l} C_i$. For every partition $\pi$ of $C$ into the
two subsets $A$ and $B$, we denote 
${\cal R}(\pi) = \{V \in {\cal F} : |V \cap A| > k \}$,
${\cal L}(\pi) = \{ V \in {\cal F} : |V \cap B| > l \}$,
and ${\cal M}(\pi)= \{ V \in {\cal F} : |V \cap A|=k~~and~~|V \cap B|=l \}$.  
We now start with the partition of $C$ into the two subsets $A=C$ and 
$B=\emptyset$, and start moving one element at a time from $A$ to $B$ 
until we obtain a partition $\pi_1$ of $C$ into the two subsets $A$ and $B$ 
and a partition $\pi_2$ into the two subsets $A'=A-\{c\}$ and $B'=B \cup \{c\}$,
such that $|{\cal R}(\pi_1)| > k$ and $|{\cal R}(\pi_2)| \leq k$.  
It is easy to that 
${\cal L}(\pi_2) \subseteq {\cal L}(\pi_1) \cup {\cal M}(\pi_1)$,    
and therefore $|{\cal L}(\pi_2)| < l$. 
We now partition ${\cal M}(\pi_2)$ into two subsets ${\cal M}_1$ and 
${\cal M}_2$, such that ${\cal F}_1 = {\cal R}(\pi_2) \cup {\cal M}_1$
has size $k$ and ${\cal F}_2 = {\cal L}(\pi_2) \cup {\cal M}_2$
has size $l$. 
For every set $S \in {\cal F}_1$ we choose a subset 
$S' \subseteq S \cap A'$ of size $k$, whereas for every $T \in {\cal F}_2$  
we choose a subset $T' \subseteq T \cap B'$ of size $l$.
Since $A'$ and $B'$ are disjoint, we have that $S' \cap T' = \emptyset$
for every $S \in {\cal F}_1$ and $T \in {\cal F}_2$. $\Box$
\begin{lemma}
\label{l72}
Suppose that $k,m \geq 1$. If ${\cal F}$ is a family of $km$ sets
of size $km$, then it is possible to partition ${\cal F}$ into the  
$m$ subsets ${\cal F}_1, \ldots , {\cal F}_m$, each of
size $k$, and to choose for each set $S \in {\cal F}$ a subset 
$S' \subseteq S$ of size $k$, so that $S' \cap T' = \emptyset$ for
every $i \neq j$, $S \in {\cal F}_i$ and $T \in {\cal F}_j$.
\end{lemma}
{\bf Proof}\,
By induction on $m$. For $m=1$ the result is trivial. Assuming that
the result is true for $m$, $m \geq 1$, we prove it is true for $m+1$.
Let ${\cal F}$ be a family of $k(m+1)$ sets of size $k(m+1)$. 
By lemma~\ref{l71}, it is possible to partition ${\cal F}$ into
a family ${\cal F}_1$ of $k$ sets and a family ${\cal F}_2$ of $km$ sets,
to choose for each $S \in {\cal F}_1$ a subset $S' \subseteq S$ of size $k$,
and to choose for each set $T \in {\cal F}_2$ a subset $T' \subseteq T$
of size $km$, so that $S' \cap T' = \emptyset$ for every $S \in {\cal F}_1$
and $T \in {\cal F}_2$. The proof is completed by applying the induction
hypothesis on ${\cal F}_2$. $\Box$

\noindent
{\bf Proof of Theorem~\ref{t72}}\,
Let $G=(V,E)$ be a strongly $k$-choosable graph. Let $V_1, \ldots ,V_r$ be
pairwise disjoint subsets of $V$, each of size at most $km$. Let $H$ 
be the graph $[G,V_1, \ldots ,V_r]$. To complete the proof, it suffices to 
show that $H$ is $km$-choosable. For each $v \in V$, let $S(v)$ be a set of 
size $km$. By lemma~\ref{l72}, for every $i$, $1 \leq i \leq r$, is it
possible to partition $V_i$ into the $m$ subsets $V_{i,1}, \ldots ,V_{i,m}$,
each of size at most $k$, and to choose for each vertex $v \in V_i$
a subset $C(v) \subseteq S(v)$ of size $k$, so that 
$C(u) \cap C(v) = \emptyset$ for every $p \neq q$,
$u \in V_{i,p}$ and $v \in V_{i,q}$.
Since the graph $[G,V_{1,1}, \ldots ,V_{r,m}]$ is $k$-choosable, we can 
obtain a proper vertex-coloring of $H$ assigning to each vertex a color 
from its set. $\Box$ 

\noindent
{\bf Proof of Theorem~\ref{t73}}\,
Apply lemma~\ref{l72} as in proof of theorem~\ref{t72}. $\Box$

\noindent
{\bf Proof of Corollary~\ref{c71}}\,
If is proved in~\cite{FS} that if $G$ is a $4$-regular graph on $3n$ vertices
and $G$ has a decomposition into a Hamiltonian circuit and $n$ pairwise
vertex disjoint triangles, then $ch(G)=3$. The result follows from
theorem~\ref{t73}. $\Box$

\noindent
{\bf Proof of Theorem~\ref{t74}}\,
Since $s\chi(1)=2$, we can assume that $d>1$. Suppose first that $d$ is even,
and denote $d=2r$. Construct a graph $G$ with $12r-3$ vertices,
partitioned into $8$ classes, as follows. Let these classes
be $A,B_1,B_2,C_1,C_2,D_1,D_2,E$, where $|A|=|D_1|=|D_2|=2r$, $|B_1|=|B_2|=r$,
$|C_1|=|C_2|=r-1$, and $|E|=2r-1$. Each vertex in $A$ is joined by edges to
each member of $B_1$ and each member of $B_2$. Each member of $D_1$ is 
adjacent to each member of $D_2$. Consider the following partition of the
set of vertices of $G$ into three classes of cardinality $4r-1$ each:
$$
V_1=B_1 \cup C_1 \cup D_1, V_2=B_2 \cup C_2 \cup D_2, V_3=A \cup E.
$$

We claim that $H=[G,V_1,V_2,V_3]$ is not $(4r-1)$-colorable. In a proper
$(4r-1)$-vertex coloring of $H$, every color used for coloring the vertices 
of $A$ must appear on a vertex of $C_1 \cup D_1$ and on a vertex of 
$C_2 \cup D_2$. Since $|C_1 \cup C_2| < |A|$, there is a color used for 
coloring the vertices of $A$ which appears on both $D_1$ and $D_2$. 
But this is impossible as each vertex in $D_1$ is adjacent to 
each member of $D_2$. Thus $s\chi(G) > 4r-1$ and as the maximum degree 
in $G$ is $2r$, this shows that $s\chi(2r) \geq 4r$. 

Suppose next that $d$ is odd, and denote $d=2r+1$. Construct a graph $G$ 
with $12r+3$ vertices, partitioned into $8$ classes, as follows. Let these
classes be named as before, where $|A|=|D_1|=|D_2|=2r+1$, $|B_1|=r+1$,
$|C_1|=r-1$, $|B_2|=|C_2|=r$, and $|E|=2r$. In the same manner we can prove
that $[G,V_1,V_2,V_3]$ is not $(4r+1)$-colorable. Thus $s\chi(G) > 4r+1$ 
and as the maximum degree in $G$ is $2r+1$, this shows that
$s\chi(2r+1) \geq 4r+2$, completing the proof. $\Box$

\end{document}